# Planetary Nebula NGC 2818: Revealing its complex 3D morphology

S. Derlopa[1]★, S. Akras[1]★, P. Amram,[2] P. Boumis[1]★, A. Chiotellis[1] and C. Mendes de Oliveira[3]

[1]*Institute for Astronomy, Astrophysics, Space Applications and Remote Sensing, National Observatory of Athens, 15236 Penteli, Athens, Greece*
[2]*Aix-Marseille University, , CNRS, CNES, LAM, Marseille, France, UFR Sciences, Physics department, 13013 Marseille*
[3]*Instituto de Astronomia, Geofísica e Ciências Atmosféricas da U. de São Paulo, Cidade Universitária, 05508-900 São Paulo, SP, Brazil*

**ABSTRACT**
We carry out an advanced morpho-kinematic analysis of the Planetary Nebula (PN) NGC 2818, whose complex morphology is described by a basic bipolar component, filamentary structures and a knotty central region. We performed an upgrated 3D Morpho-kinematic (MK) model by employing the SHAPE software, combining for the first time in PNe optical 2D spatially resolved echelle spectra and Fabry–Perot data cubes. The best-fitting 3D model of NGC 2818 successfully reconstructs the main morphology, considering one bipolar component, radial filamentary structures, and an equatorial component as the geometrical locus of the group of cometary knots. The model shows that the equatorial component has the lower expansion velocity of the system at $70 \pm 20\,\text{km s}^{-1}$. The velocity of the bipolar component is $120 \pm 20\,\text{km s}^{-1}$, while all the filamentary structures were found to expand at higher velocities of $180 \pm 20\,\text{km s}^{-1}$. Moreover, Fabry–Perot data revealed for the first time a north-eastern filament expanding at a mean velocity of $80 \pm 20\,\text{km s}^{-1}$, while its equivalent counterpart in the south-western region was confirmed by a new detected substructure in the echelle data. A new detected knotty structure at velocity $-40 \pm 20\,\text{km s}^{-1}$ is also reported, as expelled material from the fragmented eastern lobe of the nebula. We interpret the overall structure of NGC 2818 as the result of the evolution of a binary system that underwent the common envelope phase, in conjunction with the ejections of a magnetized jet, misaligned with respect to the symmetry axis of the bipolar/elliptical shell.

## 1 INTRODUCTION

Planetary Nebulae (PNe) represent the final evolutionary stage of low and intermediate mass stars ($\sim$1–8 M$_\odot$). They consist of multiple expanding shells ejected by the episodic mass outflows of the progenitor star towards the formation of a white dwarf (WD). The circumstellar envelope that has been formed due to the ejections is subsequently illuminated and reshaped by both the ionizing (ultraviolet – UV) radiation and the fast winds of the exposed hot WD, respectively (Shklovsky 1956; Kwok 2000). Given the large variety of shapes that PNe display which are very often accompanied by (micro)structures such as filaments, arcs, inner bubbles of glowing gas, knots and jet-like structures, (e.g. Derlopa et al. 2019; Guerrero et al. 2021; Mari; Gonçalves & Akras 2023; Moraga Baez et al. 2023), it is widely accepted that their formation process is not as simple as ejected shells glowing by a hot stellar remnant. Furthermore, the recently published astonishing *JWST* images unveiled PNe structures of incredible detail, confirming the intricate evolutionary history that PNe may have (e.g. De Marco et al. 2022; Wesson et al. 2023).

With respect to the spherically symmetric PNe, the formation mechanism that is currently widely accepted was proposed by Kwok, Purton & Fitzgerald (1978) known as *Interacting Stellar Winds* (*ISW*) model. According to the ISW model a slow ($\sim$10 km s$^{-1}$) and dense stellar wind which had been ejected during the asymptotic giant branch (AGB) phase, is swept up by a fast ($\sim$1000 km s$^{-1}$) and tenuous stellar wind ejected from the post-AGB star, resulting to the creation of a compressed shell. Interior to this dense rim there is a tenuous and very hot bubble ($>10^6$ K, X-ray emission) which surrounds the central star (e.g. Ruiz et al. 2011; Toalá et al. 2020 and references therein). Although the ISW scenario accounts for spherically symmetric PNe, it is not adequate for more complex cases, such as the bipolar PNe.

Bipolar PNe is a subclass of particular scientific interest, also known as 'butterfly' nebulae (Balick 1987; Dayal et al. 2000; Balick et al. 2018; de la Fuente et al. 2022; Kastner et al. 2022). In this subclass, we encounter a pair of bipolar lobes – usually of equal lengths – aligned with each other on the same axis, and a 'waist' in the middle of the lobes, which is usually denser and brighter than the extended lobes. According to Sahai, Morris & Villar (2011), bipolar PNe count to nearly 30 per cent of PNe, based on a sample of 119 PNe. This number represents only a lower limit, given that seemingly round PNe can actually be bipolar, viewed along the major axis (e.g. Meaburn et al. 2005).

In an attempt to justify bipolarity in PNe, Kahn & West (1985) and Balick (1987) suggested that bipolarity is attributed to the presence of a high pole-to-equator density contrast in the AGB slow wind.

★ E-mail: sophia.derlopa@noa.gr (SD); stavrosakras@noa.gr (SA); ptb@astro.noa.gr (PB)

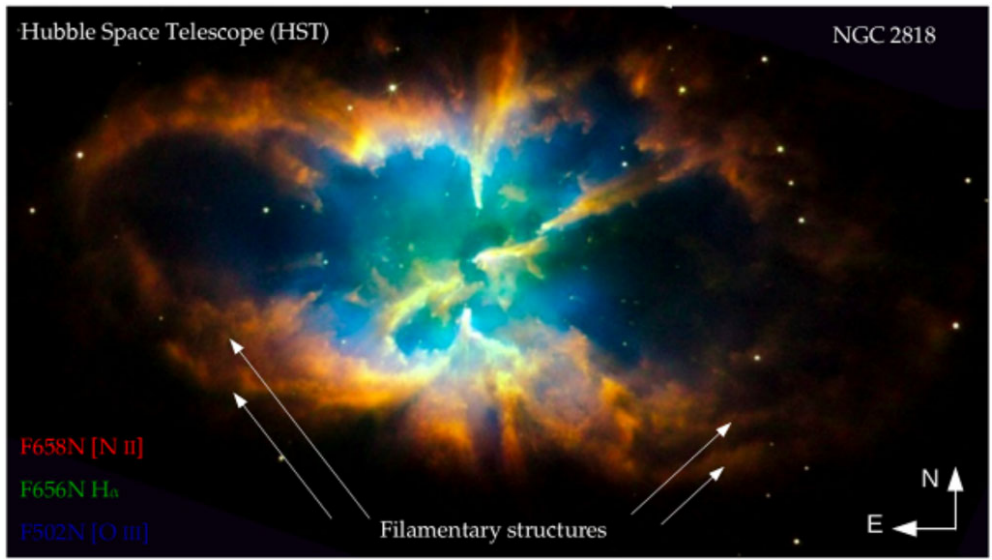

**Figure 1.** Composite image of the PN NGC 2818 from the *Hubble Space Telescope* (*HST*), in H α, [N II], and [O III] emission lines (proposal ID: 11956, PI: Keith Noll). Characteristic attributes of the nebula are the bulk of cometary knots in its central region, as well as the fragmentation at the edges of both lobes. The arrows indicate the filamentary structures which are present mostly at the outer edges of the lobes.

This model, which was an extension of the ISW proposed scenario, became known as the *Generalized Interacting Stellar Winds* (*GISW*) model (for review, see Balick & Frank 2002).

The development of a dense equatorial region in a bipolar PN might be associated with the presence of a close-binary system evolving through their common envelope (Paczynski 1976, see also the review paper of Jones & Boffin 2017 and references therein). Moreover, binary interaction can also result in the formation of jets/collimated fast winds perpendicular to the orbital plane of the binary system or the formed accretion disc, forming by this way the bipolarity we witness in many PNe (Soker & Rappaport 2000, and Soker 2002 and references therein). On the other hand, García-Segura & López (2000) showed that binarity is not absolutely necessary for bipolarity. They showed that the formation of bipolar PNe can be attributed to the misalignment between the magnetic collimation axis with respect to the symmetry axis of the bipolar/elliptical shell. Depending on the range of this angle, the additional feature of fragmentation can also be present in PNe.

In the category of bipolar PNe falls NGC 2818, the study case of this paper, which has been classified as a typical bipolar nebula (Fig. 1). The first 3D Morpho-kinematic (MK) study of NGC 2818, by employing the SHAPE software (Steffen et al. 2011), was carried out by Vázquez (2012) considering a simply bipolar structure. However, the nebula displays a more complex structure with open-ended or broken lobes, multiple filamentary structures, and several cometary knots, the latter mostly centrally concentrated. Moreover, the double filaments observed in the southern part of the nebula (marked with arrows in Fig. 1) may imply the intersection of more than one pairs of lobes. This complexity motivated us to revisit NGC 2818 under the prism of its 3D MK structure, also by employing the SHAPE software, and taking the advantage of combining high-resolution echelle long-slit spectra and high-resolution Fabry–Perot (FP) data. Through this upgrated 3D model, we aim to shed light on the inner structures of NGC 2818 which in turn will help us to further clarify its evolutionary history that resulted in its current morphology.

The paper is organized as follows: Section 2 presents the characteristics of NGC 2818, in Section 3 the observational part is presented, Section 4 presents the Fabry–Perot data for NGC 2818, and Section 5 describes the SHAPE modelling of the nebula. The results of this work are discussed in Section 6, and we end up with the conclusions in Section 7.

## 2 THE PLANETARY NEBULA NGC 2818

NGC 2818 (PK261.9 + 08.5, $\alpha_{2000}$: $09^h16^m01.656^s$, $\delta_{2000}$: $-36°37'38.76''$) displays a bipolar morphology with a knotty central region and filamentary structures up to its outer edges (see Fig. 1 and Vázquez 2012). The nebula has been assumed to be a member of an open cluster of the same designation (Tifft, Connolly & Webb 1972), but more recently Mermilliod et al. (2001) argued that the membership of the nebula is very unlike due to the inconsistency between the radial velocity of the nebula ($-1$ km s$^{-1}$, Meatheringham, Wood & Faulkner 1988) and the cluster's mean velocity (20.69 km s$^{-1}$). However, our mean radial (or systemic) velocity of NGC 2818 was found to be $23 \pm 4$ km s$^{-1}$ (see below Section 3.2) which supports a membership of the nebula.

Distance estimates of NGC 2818 range from 1.79 up to 3.5 kpc: 1.79 kpc (Phillips 2004); 2.3 kpc (van de Steene & Zijlstra 1995); 3.05 kpc (Zhang 1995); and 3.5 kpc (Dufour 1984). GAIA parallaxes (DR2: $0.2712 \pm 0.3216$ mas; Gaia Collaboration 2018, DR3: $0.0319 \pm 0.2111$ mas; Gaia Collaboration 2021) are not ideal to compute the distance of the nebula just by inverting them because of the large uncertainties. The geometric distances of the nebula based on the GAIA parallaxes are 2.39 (lower upper bound, 1.49–4.01) kpc (Bailer-Jones et al. 2018) and 4.17 (lower upper bound, 2.91–5.88) kpc (Bailer-Jones et al. 2021), respectively.[1] The former distance is within the range of the previous independent estimates, while the latter places the nebula even further away.

---

[1]These distances are computed correcting for the zero-point in DR2 ($-0.029$ mas) and DR3 ($-0.017$). However, there is no proper determination of the zero-point for hot, blue stars as the PNe central star and this correction can be larger (Gómez-Gordillo et al. 2020).

Regarding the chemical composition, the nebula is characterized by high He and N abundances, classified as Type I (Peimbert & Torres-Peimbert 1983; Dufour 1984). The latter is also supported by the bright equatorial region which is fragmented in numerous knots and filaments and a pair of open-ended lobes that the nebula displays (Fig. 1), which is also characteristic of Type I PNe. The mass of the nebula has been computed nearly at $0.6 \pm 0.2$ M$_\odot$, while the mass of the progenitor star is $2.2 \pm 0.3$ M$_\odot$ (Dufour 1984).

Concerning its radiation, NGC 2818 mostly emits in the infrared (IR) and in the optical band. With respect to the IR emission, Schild (1995) reported the detection of the H$_2$ ($\lambda = 2.12\,\mu$m) co-vibrational line. This emission displays a seemingly ellipsoidal shell, that appears co-spatial with the H$\alpha$ emission in the central and eastern part of the nebula. Furthermore, the investigation of NGC 2818 in the mid-infrared (MIR) with *Spitzer* showed that its morphology in the 3.6, 4.5, 5.8, and 8 $\mu$m band-passes follows the general structure observed in the optical and H$_2$ emission (Phillips & Ramos-Larios 2010). An increasing surface brightness towards longer MIR wavelengths (5.8 and 8.0 $\mu$m) has also been found, which may be associated with the emission of H$_2$ lines, with forbidden atomic lines and/or warm dust (Hora et al. 2004).

In general, the increase of the emission in the 4.5 $\mu$m band has been attributed to a potential contribution of H$_2$ emission, as it has been observed in the cometary knots of Helix (Hora et al. 2006, Matsuura et al. 2009, Meaburn & Boumis 2010) and the low-ionization structures (LIS; e.g. Gonçalves, Corradi & Mampaso 2001; Akras & Gonçalves 2016) of NGC 7009 (Akras et al. 2020). The low value of the 8.0/4.5 $\mu$m ratio observed in the centre of NGC 2818 is consistent with the detection of strong H$_2$ emission (Schild 1995). Additional study in the MIR spectrum of NGC 2818 with *Spitzer* has demonstrated the presence of several molecular lines from different vibration levels, as well as high excitation forbidden lines (Mata et al. 2016). The analysis of H$_2$ spectrum resulted in a rotational excitation temperature close to $850 \pm 50$ K without any conclusion about the excitation mechanisms of H$_2$.

Regarding the optical emission, NGC 2818 displays strong emission lines in H$\alpha$, [N II], and [S II], and more centrally concentrated emission in [O III] 5007 Å and He II 4686 Å lines (Dufour 1984; Phillips & Cuesta 1998; Górny et al. 1999; Vázquez 2012). The spectroscopic study of NGC 2818 by Phillips & Cuesta (1998) revealed a significant electron density fluctuations based on the [S II] diagnostic lines, supporting the scenario of dense clumps/filaments embedded in a lower density environment. The same authors also reported that the high [S II]/H$\alpha$ and [N II]/H$\alpha$ line ratios found in NGC 2818 in conjunction with the strong H$_2$ emission detected in the southern part of the nebula, are attributed to shock interactions. The link between strong low-ionization lines and H$_2$ emission has also been reported in the small-scale LISs of PNe (Akras, Gonçalves & Ramos-Larios 2017; Akras et al. 2020).

In the most recent study of NGC 2818, Vázquez (2012) confirmed the distribution of a great number of cometary knots in the nebula using high-resolution optical images from *HST*. The same author also presented the first 3D MK model of this nebula, by employing the astronomical software SHAPE (Steffen et al. 2011). This particular model was constructed considering a simple bipolar shell. However, the overall complexity in its structure was our motivation in order to develop a more detailed 3D MK model. For this study, we combine high-quality *HST* images, two-dimensional echelle spectra and three-dimensional Fabry–Perot spectroscopic data. The combination of these sets of data will help us to constrain even further the structure of the nebula and its evolutionary history.

# 3 OBSERVATIONS

## 3.1 High resolution imaging

High spatial-resolution *HST* images of NGC 2818 were drawn from the Hubble Legacy Archive.[2] The observations were carried out in 2008 November 27$^{th}$ (proposal ID: 11956, PI: Keith Noll). The Wide Field Planetary Camera 2 (WFPC2) with $800 \times 800$ pixels$^2$, image scale of 0.2 arcsec pixel$^{-1}$ and Field of View (FOV) of $2.7 \times 2.7$ arcmin$^2$ was used.

The narrow-band F469N, F502N, F656N, F658N, and F673N filters were used to isolate the He II $\lambda$4686, [O III] $\lambda$5007, H$\alpha$ $\lambda$6563, [N II] $\lambda$6584, and [S II] $\lambda$6731 emission lines and determine the morphology of the nebula in different ionization states. The exposure times for the above emission lines were 2000, 2000, 1600, 2000, and 2000 s, respectively. All the *HST* images are presented in Vázquez (2012). The RGB colour image of NGC 2818, mapping [N II] to red, H$\alpha$ to green, and [O III] to blue, is presented in Fig. 1 (proposal ID: 11956, PI: Keith Noll).

## 3.2 High resolution echelle spectroscopy

Fig. 2 (left image) displays the slit positions overlaid on the *HST* [N II] $\lambda$6584 image of NGC 2818. The spectroscopic data obtained from slits 1 to 5 were retrieved from the work of Vázquez 2012 (private communication) from observations which were carried out in 2009 February 4–6, while the spectrum from slit 6 which was obtained in 2007 February 4th, was downloaded from the SPM Kinematic catalogue of Galactic PNe (López et al. 2012). All the echelle spectra were obtained using the Manchester Echelle Spectrometer (MES-SPM; Meaburn et al. 2003) on the 2.1 m telescope in its *f*/7.5 configuration at the San Pedro Martir Observatory in Baja California (Mexico).

The H$\alpha$ $\lambda$6563 Å and [N II] $\lambda\lambda$6548, 6584 Å nebular emission lines were isolated by the 87$^{th}$ echelle order using a 90 Å bandwidth filter. The SITE3 CCD detector with $1024 \times 1024$ pixels$^2$, each of 24 $\mu$m was used. A $2 \times 2$ binning was also considered in both spatial and spectral directions in order to increase the signal-to-noise ratio of the emission lines, resulting in a spatial scale of 0.624 arcsec pixel$^{-1}$. The 150 $\mu$m-width slit or 1.9 arcsec wide on the sky was used, resulting in a spectral resolution of 11.5 km s$^{-1}$. The position angle (hereafter P.A.) of slits 1, 2, and 6 was 89°, 123°, and 89°, respectively. Slits 3, 4, and 5 have P.A. = $-1$°. The Th/Ar calibration lamp was used for the wavelength calibration of the data. The data reduction (bias correction, flat-fielding, and cosmic rays cleaning) was performed using the standard IRAF routines. All spectra were calibrated to radial velocity, which is the Doppler velocity after the heliocentric and systemic correction ($23 \pm 4$ km s$^{-1}$). The deduced Position–Velocity (PV) diagrams in the [N II] $\lambda$6584 line are shown in the right panels of Fig. 2.

The [N II] $\lambda$6584 emission line was specifically used for the reconstruction of the 3D MK model of NGC 2818, because the low thermal broadening that presents results in the depiction of more sharp structures of the nebula.

## 3.3 Fabry–Perot high resolution spectroscopy

Fabry–Perot (FP) observations were taken at the 4.1 m Southern Astrophysical Research (SOAR) Telescope, located at Cerro Pachón,

---
[2] https://hla.stsci.edu/hlaview.html

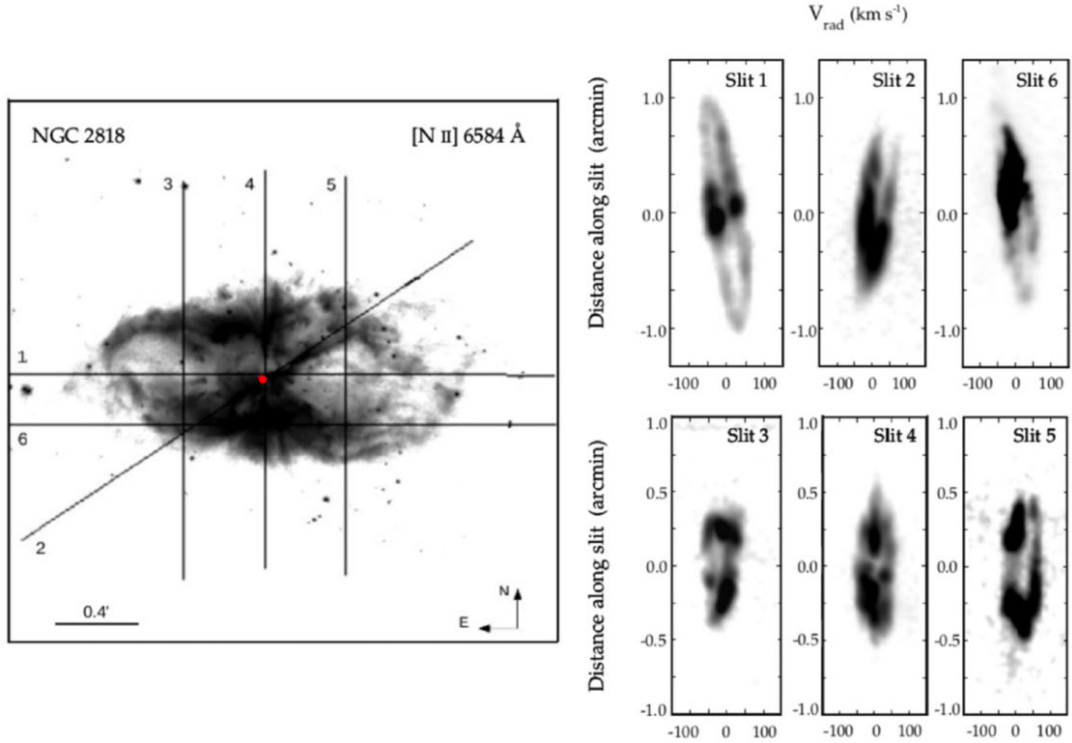

**Figure 2.** *Left image*: the *HST* image of NGC 2818 in [N II] 6584 Å emission line. The overlaid numbered lines represent the positions of the six slits where echelle spectra were obtained. The dot marks the position of the central star. *Right image*: the six Position–Velocity (PV) diagrams in the [N II] 6584 Å emission line deduced from the echelle data. The horizontal axis is calibrated in $V_{\rm rad}$ with range $(-150, +150)\,{\rm km\,s^{-1}}$, while the vertical axis is calibrated in arcmin (distance along the slit).

Chile. The SOAR Adaptive Module-Fabry–Perot (SAM-FP) is a restricted-use instrument which consists of a scanning FP device mounted inside the SOAR Adaptive Module (SAM, Tokovinin et al. 2008). The SAM-FP instrument has been described in Mendes de Oliveira et al. (2017). Under photometric conditions, H$\alpha$ and [N II] $\lambda$6584 emission lines data cubes of NGC 2818 were obtained on UT date 2017, in March 2 and April 1, respectively, during two different observing runs. We made 3600 s of exposures on the H$\alpha$ line and 2880 s on the [N II] $\lambda$6584 line, plus calibrations. The seeing during both sets of observations varied between 0.7 and 0.9 arcsec. The ground-layer adaptive optics (GLAO) of SAM improves the spatial resolution data cubes in concentrating the flux of up to 2. Nevertheless, the GLAO did not work when we observed the H$\alpha$ data cube, but yielded to a FWHM of 0.6–0.7 arcsec, determined by the laser-corrected FWHM, during the observation of the [N II] line. An e2v CCD with 4096 × 4112 pixels$^2$ provided a FOV of 3 × 3 arcmin$^2$ and a pixel size (binned 4 × 4) of 0.18 × 0.18 arcsec$^2$. Circular 3 inches Cerro Tololo interference filters centred on 6569 and 6584 Å, with a FWHM of 20 Å and a transmission of ∼70 per cent for both, were placed in the focal plane to select the wavelength ranges. An ICOS ET-652 Fabry–Perot interferometer, working at an interference order $p = 609$ at H$\alpha$ that provides a free spectral range FSR = 492 km s$^{-1}$ was covered in 48 scanning steps of 10.3 km s$^{-1}$, for both lines. The effective finesse, i.e. the instrumental profile, as measured from Ne calibration lamp lines, was $F = 18.5$, offering a resolving power RP ∼ 11.266 at H$\alpha$, i.e. a spectral resolution of 0.583 Å or 26.6 km s$^{-1}$. The error of the instrument is of the order of 15–20 km s$^{-1}$.

Raw images obtained with SAM-FP were transformed into scientifically useful data cubes after following several procedures described in Mendes de Oliveira et al. (2017). Briefly, we estimated the background using the overscan region by fitting and subtracting a third degree polynomial function. We subtracted the bias and darks from the science frames and corrected them by the normalized flats to remove the instrumental signature. The wavelength calibration was performed, using the narrow Ne I line (6598.95 Å), through the building of a phase map that provides the same wavelength origin to each spaxel. The phase map was thus applied to the raw data cube to construct the wavelength calibrated data cube where each frame represents one wavelength, modulo the FSR, given the cyclic nature of the Fabry–Perot interferometer. From the wavelength calibrated cube, we calculated the moments of order zero (line flux), one (radial velocities), and two (dispersion velocities). This has been done in measuring the barycenters and the widths of the emission lines, spaxel per spaxel. The velocity dispersion maps ($\sigma$) were corrected for the instrumental response by quadratically subtracting the LSF (Line Spread Function) from the measured widths, supposing that all the functions could be estimated by Gaussians following $[\sigma = (\sigma_{\rm LoS} - \sigma_{\rm LSF})(1/2)]$.

## 4 SAM FABRY–PEROT DATA OF NGC 2818

The angular size of NGC 2818 (2 arcmin × 1 arcmin) fits perfectly in the FOV of the SOAR SAM-FP instrument (see Section 3.3), which makes the nebula an ideal target for a thorough morpho-kinematic study covering all of its extent. For the first time, maps which provide detailed kinematic information of the whole nebula are presented, as these were derived from the 2D collapsed SAM-FP data cubes. These maps, which are described in Sections 4.1 and 4.2 below, confirm in the most clear way the complexity of the kinematics of NGC 2818, something that had already been denoted by the previously published long slit spectra (Vázquez 2012).

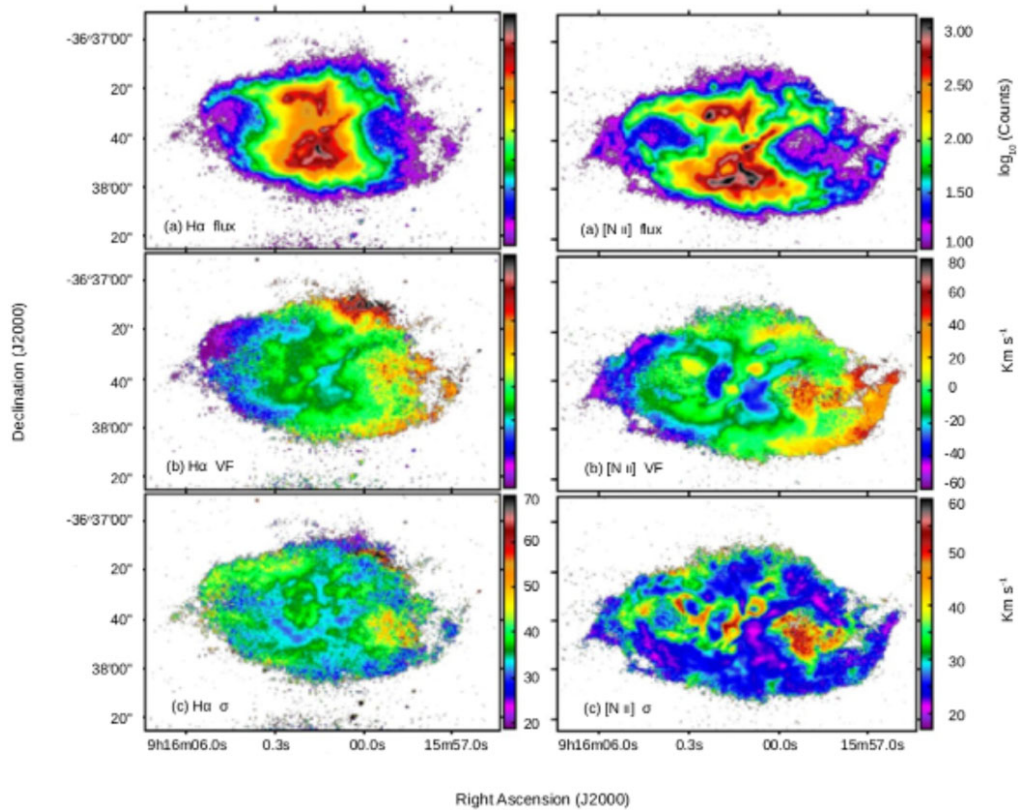

**Figure 3.** On the left: H α monochromatic (top panel), velocity field (VF, middle panel), and dispersion velocity maps (σ, bottom panel) of NGC 2818. H α monochromatic map is not flux calibrated, and the colour bar indicates the counts per pixel in logarithmic scale. Velocity field and dispersion velocity maps are corrected by the instrumental and thermal broadening. The scale is shown in km s$^{-1}$. On the right: the same for the [N II] 6584 Å line.

### 4.1 H α and [N II] λ6584 monochromatic maps

In Fig. 3(a), we present the monochromatic maps of NGC 2818 for the H α (on the left) and [N II] λ6584 (on the right) emission lines, respectively.

It can be seen that the H α emission is less extended than the [N II] λ6584 emission by a factor of ∼10 per cent. For both emission lines, the inner part of the nebula is characterized by a factor of ∼1000 higher surface brightness than the outer regions (see equivalent colour bars). The lowest surface brightness can be seen in the two inner cavities (flux < 1.50 in log scale), one on either side of the nebula's minor axis. In contrast, two bright areas are observed one on each side of the major axis (flux > 2.25 in log scale). As well for the cavities as for the bright regions, a greater density contrast is observed in [N II] λ6584 than in the H α. The combination of these two structures (cavities and brighter areas) results in an asymmetrical X-shape, brighter in the south than in the north, and bright both in the east and in the west, which gives the morphology of an irregular bipolar nebula.

The outer part of the nebula forms an elliptical rim made of, more or less, bright arc segments in the east–west direction. The P.A. of the major-axis is nearly to 90 degrees and the derived ellipticity is b/a ∼ 0.53 ± 0.03 and ∼0.55 ± 0.03 for the H α and [N II] λ6584 maps, respectively. These values do not vary much with the thresholds to delineate the ellipse. The actual angular size of NGC 2818 in the H α ([N II] λ6584) line is computed 120 (160) arcsec along the major axis, and 50 (60) arcsec along the minor axis. In both emission lines, the centres of the ellipses correspond well to the bright centre of the nebula, coinciding with the position of the ionizing star. Moreover, the flux fluctuation indicates the presence of several microstructures such as knots and filaments in the central regions, in agreement with what we see in the *HST* image (Fig. 1).

### 4.2 Kinematic properties

In Fig. 3, the 2D velocity field (panels b) and velocity dispersion maps (panels c) of NGC 2818 extracted from the H α and [N II] λ6584 data cubes are presented on the left and on the right images, respectively. They are corrected by instrumental ($\sigma_{int} = 11.3$ km s$^{-1}$) and thermal broadening ($\sigma_{th} = 9.1$ km s$^{-1}$) assuming $T_e = 10000$ K in the expression $\sigma_{th} = (k\,T_e/m_H)^{\frac{1}{2}}$. Overall, the nebula displays sharp changes in velocities in small scales (<a few arcsec) which implies a high complex kinematic structure. Both velocity field and dispersion velocity maps show very little correlation with the H α and [N II] λ6584 emission (monochromatic maps – Section 4.1).

The velocity maps display the peak of a single Gaussian fit at each pixel, providing an overall view of the kinematics of the nebula. No significant spatial difference is found in the velocity field between the two emission lines. The eastern part of the nebula is blue-shifted and the western part red-shifted, with a velocity range between −60 and 80 km s$^{-1}$, which is in agreement with the velocity range of the echelle data (Fig. 2).

On the other hand, the velocity dispersion maps display the width of a single Gaussian fit at each pixel, and are connected with regions of possible multiple kinematic components. The dispersion velocity of H α line ranges from 25 to 60 km s$^{-1}$, while the range for the line [N II] λ6584 is between 20 and 55 km s$^{-1}$. However, it is in [N II] λ6584 line where both lobes show velocity dispersions up to

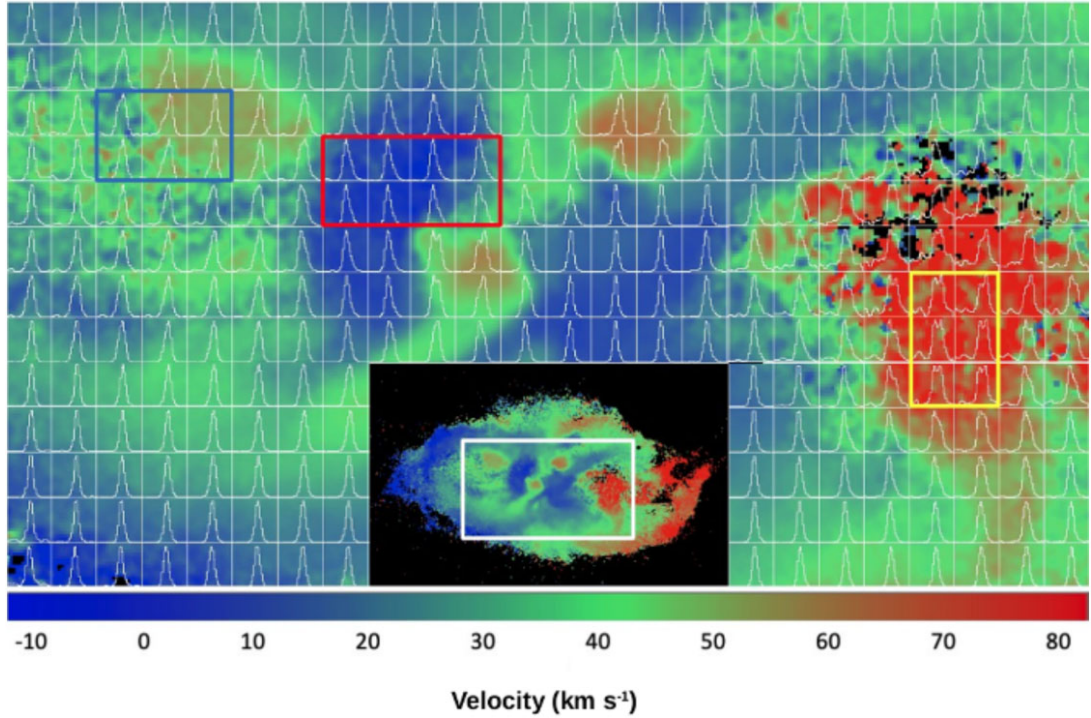

**Figure 4.** [N II] λ6584 Å emission line profiles of the central region of NGC 2818. The inset image shows the line of sight velocity for reference (similar to Fig. 3b on the right). The white rectangle in this inset is enlarged in the large figure, which shows a FOV of ∼65 × 35 arcsec$^2$, with each binned pixel to be ∼2.7 arcsec on each side; north is up and east is to the left. The profiles are not corrected from the LSF. The purpose of this figure is to show the diversity of profiles at arcsec scale . For further description see Section 4.2.

∼55 km s$^{-1}$, which indicates regions where multiple substructures are present.

The latter becomes more evident in Fig. 4, where we present the [N II] line profiles of a central region overlaid on the velocity field map. In this image, we can identify three basic kinds of line profiles: (I) the double-peak lines, like the ones included in the yellow box. They are presented in regions of high velocities (>65 km s$^{-1}$), they present comparable intensities and high dispersion, and they indicate the presence of more than one substructures; (II) asymmetric line profiles at velocities lower than 40 km s$^{-1}$, with very faint wings, marked with the blue and red boxes. They indicate the presence of unresolved components of different intensities, with blue and red-shifted parts, respectively, with respect to the main line profile; (III) symmetric single Gaussian line profiles in the rest images, out of the boxes. We can see them in regions of lower velocities too (<40 km s$^{-1}$), they illustrate low dispersion velocity and they concern a single component.

For a more detailed examination of the kinematics of NGC 2818, we display the channels maps 7–34 for the H α line (from −179 to 190 km s$^{-1}$) and for the [N II] λ6584 (from −191 to 180 km s$^{-1}$) in Figs 5 and 6, respectively, both in logarithmic scale. In particular, in both Figs 5 and 6, we clearly see the ellipsoidal shape of NGC 2818 in the velocity range from 0 to 40 km s$^{-1}$, for both emission lines. For higher velocities, the open-ended/broken lobes become evident, along with the detection of multiple filamentary components in both eastern and western lobes, which are also discerned in the *HST* image.

Regarding the central part of NGC 2818, we notice the presence of distinct regions/blobs in the northern and southern part of the nebula with velocities up to 170 km s$^{-1}$, as it is more evident in Fig. 5 (blue boxes). These high velocity components are much fainter compared to the main bipolar shell and remain undetected in this velocity range in the long-slit echelle spectroscopic data. Moreover, their velocities are surprisingly larger than these of the bipolar filaments. We argue that these components represent a bunch of high velocity knots. Overall, the SAM-FP data results, imply a more complex velocity structure than the long-slit echelle spectra provided.

### 4.3 New north-eastern filament

The spatial and spectra resolution of SAM-FP allow us to identify very faint high velocity components in NGC 2818, e.g. the knotty substructures that were mentioned in Section 4.2. In addition to that, we were also able to discover a new very faint filamentary structure which was remained undetected by the previous imaging and spectroscopic studies. This newly detected filament is visible only in the H α line in logarithmic scale between the channel maps 25 and 28 (red boxes in Fig. 5) and velocities between 67 and 108 km s$^{-1}$ (±20 km s$^{-1}$). Despite the fact that the projection of this filamentary structure is located towards the eastern side of NGC 2818, it is noteworthy that its velocity direction is red-shifted and not blue-shifted as that of the eastern lobe. For more details, see Section 6.4.

## 5 3D MORPHO-KINEMATIC (MK) MODELLING

SHAPE software (Steffen et al. 2011) has been widely used to model the 3D MK structure of PNe (e.g. Akras & López 2012; Akras & Steffen 2012; Vázquez 2012; Akras et al. 2015, 2016; Clyne et al. 2015; Sabin et al. 2017; Derlopa et al. 2019; Bandyopadhyay et al. 2020; Rodríguez-González et al. 2021; Danehkar 2022; De Marco et al. 2022), nova shells (e.g. Harvey et al. 2016, 2020; Kamiński et al.

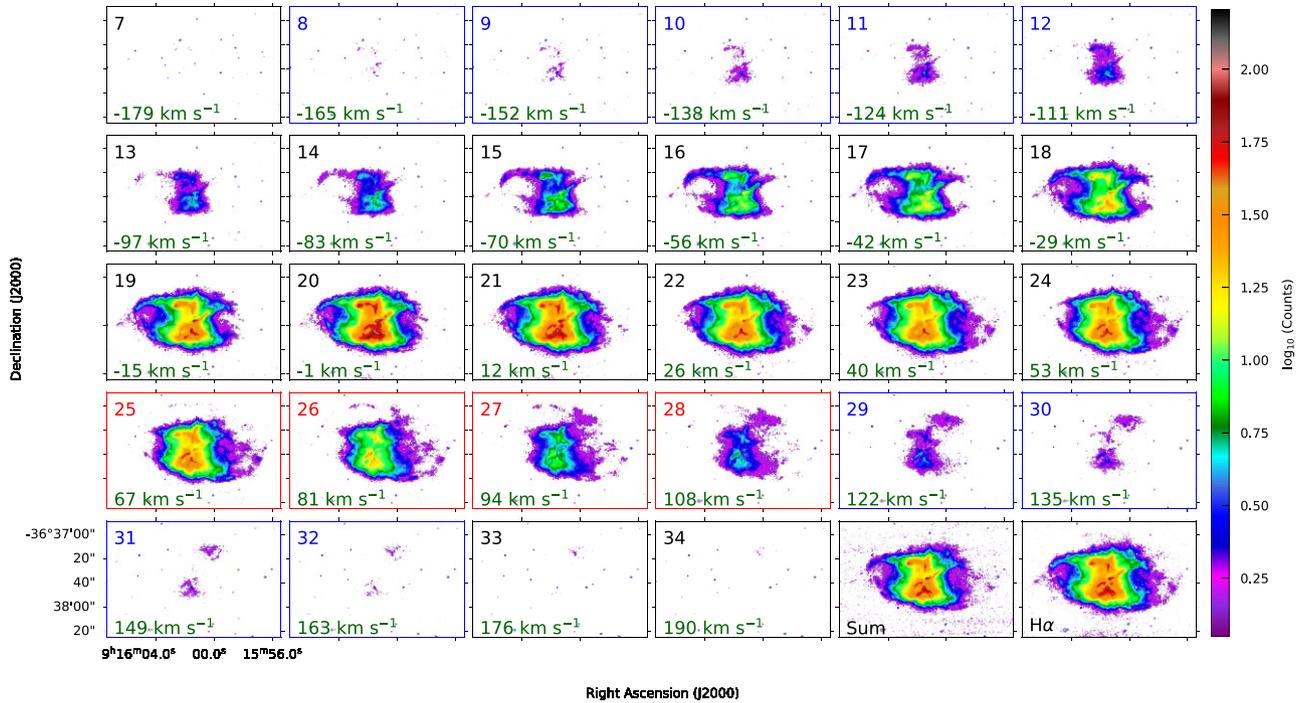

**Figure 5.** Channel maps for the H α data cube of NGC 2818 in logarithmic scale. Here are shown the 28 out of 40 channel maps of SAM-FP of most intense emission. The mean velocity of the channel and the channel number are shown at the bottom left and top left corners of each channel, respectively. The penultimate and ultimate channels show the sum of all channels and the H α emission map. The colour bar indicates the counts per pixel in logarithmic scale. In the channel maps 25–28, the new detected north-eastern filament is shown, which it is evident only in the H α line (see Sections 4.3 and 6). On the other hand, the channel maps 8-12 and 29-32 include central substructures/knots with velocities up to 170 km $^{-1}$.

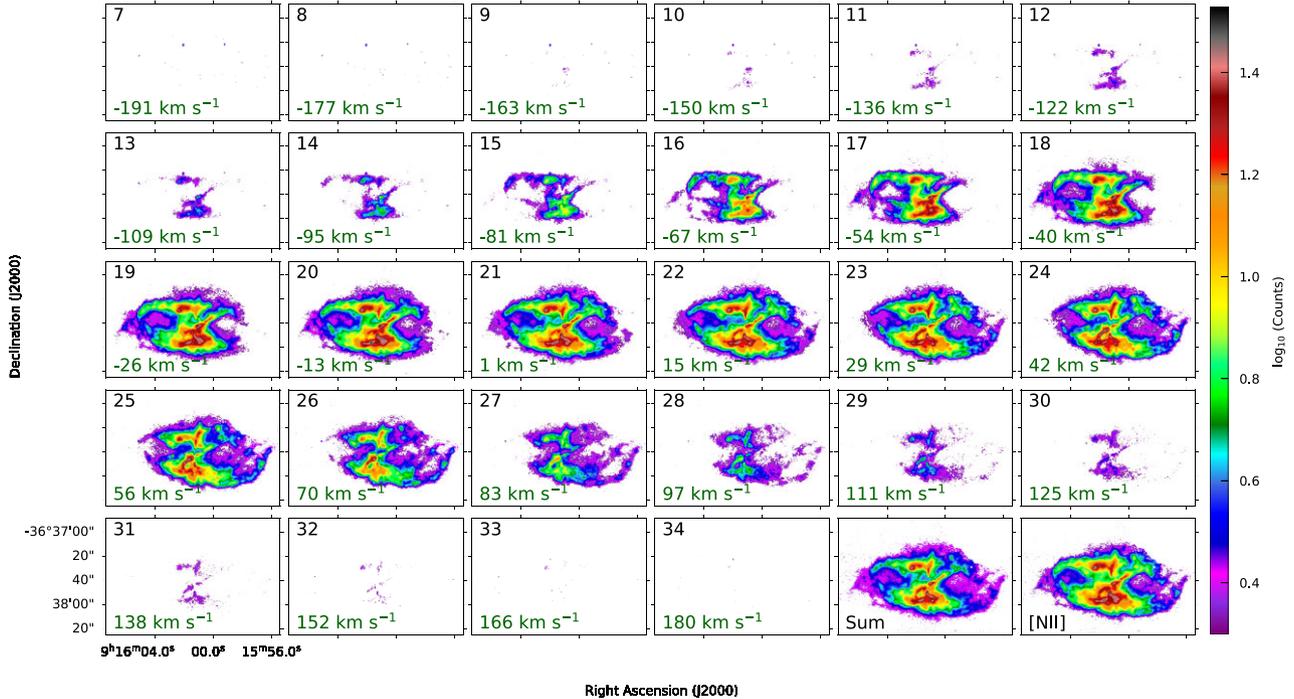

**Figure 6.** The same as Fig. 5 but for the [N II] λ6584 emission line.

2021) and recently the complex supernova remnant VRO 42.05.01 (Derlopa et al. 2020).

The importance of the 3D reconstruction and the projection effects in unveiling the true morphological structure of PNe is comprehensively demonstrated in the work of Chong et al. (2012), where an exa-polar nebula (i.e. a nebula consisted of three pairs of lobes, see for example Clark et al. 2013) is able to explain the shapes of several PNe just by varying the inclination and position angles of each bipolar component (hereafter BC). It is obvious that the clarification of the true morphology of a PN can be a very tricky task.

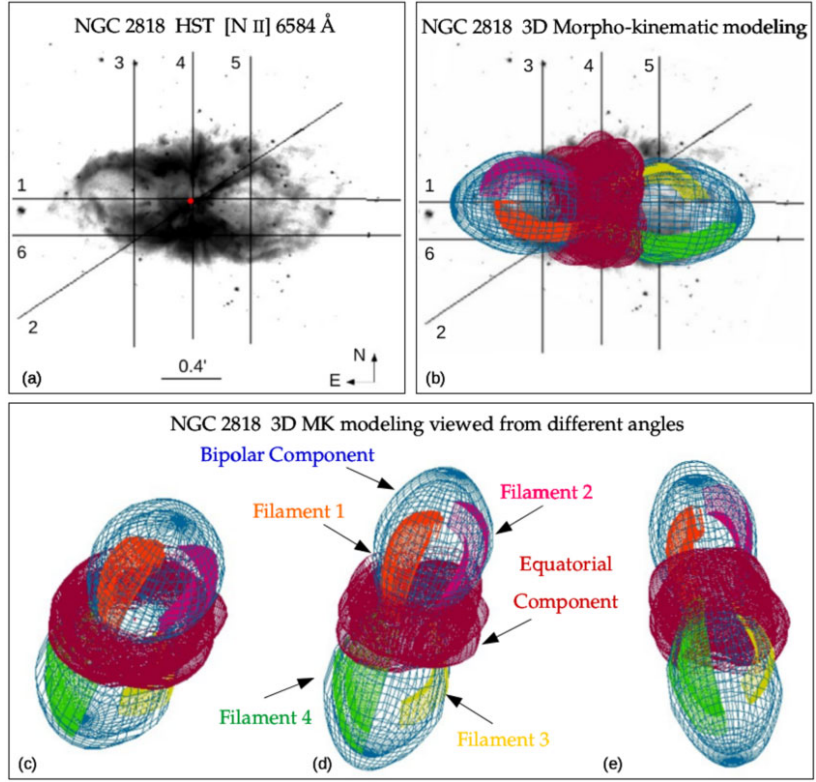

**Figure 7.** Image (a) shows the observational image of NGC 2818 from the *HST* telescope, along with the six slit positions. In image (b) the 3D MK modelling of the PN is illustrated in mesh representation, overlaid on the observational image. Images (c), (d), and (e) present the 3D MK modelling as viewed from different angles. Also labelled are the four filaments, along with the BC and the EC ('waist'). For further description, see Section 5.

### 5.1 Modelling the nebula of NGC 2818

For the modelling of NGC 2818, we used the *HST* [N II] $\lambda$6584 emission line image as a background image in SHAPE, along with the six PV diagrams which were deduced from the equivalent echelle spectra (Fig. 2). The method was to reconstruct the 3D structure of the nebula and the synthetic PV diagrams, by comparing them with the observational imaging and spectroscopic data. Furthermore, we created the synthetic channel maps of NGC 2818, the comparison of which with the observed SAM-FP channel maps allowed us to constraint more thoroughly the 3D morphology of the nebula.

Our first attempt included more than one pair of lobes, which could reproduce reasonably well the PV diagrams, but not the channel maps as well. Therefore, we concluded that the best-fitting model which reproduces both the observational data consists of only one BC following the 3D model from (Vázquez 2012). In addition, we added filamentary structures (hereafter FIL) which were included in order to depict the bright edges of the lobes components. Finally, we considered an equatorial component (hereafter EC) to describe the knotty central region of the nebula.

The final 3D MK model of NGC 2818 in mesh representation is shown in Fig. 7 (b) overlaid on the *HST* [N II] $\lambda$6584 image. In Figs 7 (c), (d), and (e) we see the 3D model from different angles. In the following subsections, we discuss the reconstruction of each nebular component in details.

#### 5.1.1 Bipolar component

The BC of NGC 2818 is evident in Fig. 1, and it is consistent with the observational PV diagrams of Fig. 2, where we clearly see 'loops' or 'velocity ellipses' which are characteristic of radial expansion of spheroidal- and ellipsoidal- type surfaces. For the reconstruction of the BC we applied in SHAPE a linear velocity law, $V = B(\frac{r}{r_o})$. $V$ is the expansion velocity ($V_{\rm exp}$) in km s$^{-1}$ for the gas of the structure, whose projection is the Doppler (or radial) velocity that we observe with slit-spectroscopy. $B$ is the expansion velocity specifically at distance $r = r_o$ from the central star, where $r$ and $r_o$ are given in arcmin.

The main guide to constrain the bipolar structure, was the observational spectrum of slit 1 (Fig. 2). The length of the bipolar lobe was determined according to the length of the spectrum in the vertical axis of its PV, while the widths of its lobes were defined from the span of the observational 'loops' (or 'ellipses') we see in the spectrum of slit 1. The resulted structure is displayed in Fig. 7 (blue colour), while the corresponding synthetic spectrum is shown in Fig. 8 (blue colour), overlaid on the black-coloured observational spectra. The velocity of the BC was found to be $V_{\rm exp} = 120 \pm 20$ km s$^{-1}$. The BC along the major axis of the nebula, it was also demonstrated by Vázquez (2012) giving a $V_{\rm pol} = 105$ km s$^{-1}$.

Regarding its orientation, it has a P.A. of $80 \pm 3$ deg with respect to the plane of the sky (0 deg is along the north–south direction) and an inclination angle of $70 \pm 3$ deg along the line of sight (0 deg is inwards the page). The eastern part of the BC is coming towards us (blue-shifted) while its western part is moving away from us (red-shifted). In the equivalent model presented by Vázquez (2012), the P.A. is 89 deg and the inclination is 60 deg. Adopting the distance value of 4.17 kpc (Bailer-Jones et al. 2021) along with its lower and upper bounds, the kinematical age of the nebula was found to be $\tau \sim$ 9400 (lower upper bound, 6500–12300) yr. This is within the value of $8400 \pm 3400$ yr given in Vázquez (2012).

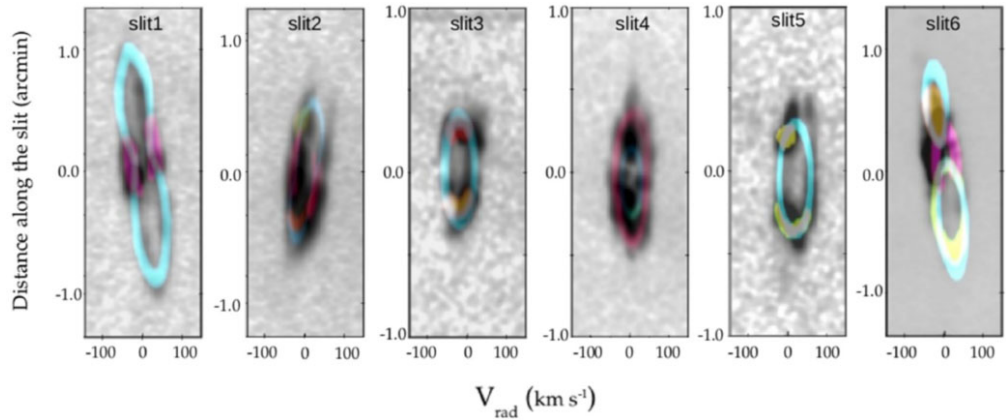

**Figure 8.** The six PV diagrams used for the 3D modelling of NGC 2818. In black are the observational spectra, while the overlaid coloured structures are the synthetic spectra reproduced with SHAPE. Each coloured structure corresponds to the spectrum resulted from the equivalent component (same colour-code as in Fig. 7) .

*5.1.2 Filamentary structures*

NGC 2818 presents a complicated filamentary formation, which is depicted on the observational data. In order to reproduce these data through our modelling, we focused on four basic extended filaments (FIL) of the nebula. Their 3D reconstruction is illustrated in Fig. 7, where each colour corresponds to a different filamentary component. Therefore, orange results from the filament FIL1, red from the filament FIL2, yellow from the filament FIL3 and green from the filament FIL4.

Each filament was reconstructed as part of the main bipolar lobe, based on the assumption that the filaments 'follow' the overall bipolar shape of the lobes, as it is shown in the *HST* image (Fig. 1), and project back to the central star. This is why they have the same P.A. = $80 \pm 3$ deg and the same inclination of $70 \pm 3$ deg as that of the BC. With respect to the velocity law of the filamentary structures, it is the same as that of the BC (Section 5.1.1). According to the model, all the filamentary structures have been expanded at higher velocities in comparison with the bipolar structure of $V_{\rm exp} = 180 \pm 20$ km s$^{-1}$. Fig. 7 illustrates the 3D mesh representation of the filaments, while in Fig. 8 the coloured structures correspond to their synthetic spectra. As we can notice, their synthetic spectra fill out the observational data at the points where the spectra present substructures of higher intensity.

*5.1.3 Equatorial component*

Of specific importance for creating a more representative model of NGC 2818 was the overall spectra of slit 4 which passes through the central region of the nebula ('waist' region) across its minor axis (north–south direction). In its synthetic spectra, only the blue part is attributed to the BC (Fig. 8). For the remaining spectrum, an additional EC was necessary to be taken into account, which is illustrated in Fig. 7.

The central region of NGC 2818 is described as the geometrical region that contains the bulk of cometary knots of the nebula. These microstructures are easily seen in the high spatial images from *HST*, but not from the ground base observation with the 2.1 m SMP telescope. This is why we were in favour of modelling the equatorial region as a toroidal structure and reproduce its overall morpho-kinematic characteristics.

The general velocity law that better constrains the kinematic characteristics of the equatorial region is also linear, and gives an expansion velocity of $V_{\rm exp} = 70 \pm 20$ km s$^{-1}$, while its P.A. and inclination are found nearly to 80 and 70 deg, respectively, in agreement with the overall direction of the rest nebula. According to the model of Vázquez (2012), the equatorial velocity is lower, $V_{\rm equat.} = 20$ km s$^{-1}$.

Regarding the synthetic spectra, the additional EC indeed fills out the observed spectrum of slit 4 (red colour in the synthetic PV, Fig. 8). Similarly, it is capable to produce the bright central regions in the PVs of slits 1, 2, and 6 (Fig. 8). Hence, including the EC in the 3D MK model of NGC 2818, we are eventually able to reproduce all the main intrinsic structures of the nebula.

*5.1.4 Channel maps with SHAPE*

Upon completion of the 3D MK model, we proceed in the extraction of the equivalent channel maps with SHAPE for the [N II] λ6584 emission line, which are illustrated in Fig. 9. We present only 12 channel maps with strong emission, from 16 to 27, where the velocity ranges from $-65$ to $85$ km s$^{-1}$ with step 14 km s$^{-1}$, equivalent to the SAM-FP channel maps of Fig. 6. For clarification reasons, we separately present the BC along with the filamentary structures in Fig. 9(a), while both components along with the EC are illustrated in Fig. 9(b). Thus, we can clearly see the contribution of each component to each channel map. The grey-scale illustration, on the right of Fig. 9, only indicates the location of each component, and does not reproduce the brightness distribution of the equivalent emission line. Therefore, we are not able to perform a channel-to-channel comparison between the SHAPE channel maps and the SAM-FP channel maps, on the basis of the flux distribution. The main goal, however, was to compare the synthetic and the observational channel maps on the basis of velocities correspondence. Indeed, the synthetic channel maps managed to reproduce the overall bipolar shape of NGC 2818, along with the characteristic open-ended lobes on both eastern–western sides of the nebula, within a velocity range which is in agreement with the observational data cubes of Fig. 6.

Furthermore, the synthetic channel maps agree with the kinematics of the nebula, on the basis that for negative velocities we only see the eastern blue-shifted lobe, while for positive velocities only the western red-shifted lobe is evident. For velocities that approach the zero velocity, the overall bipolar morphology is clear along with both lobes, while the central region of the knots locus is present in the whole velocity range.

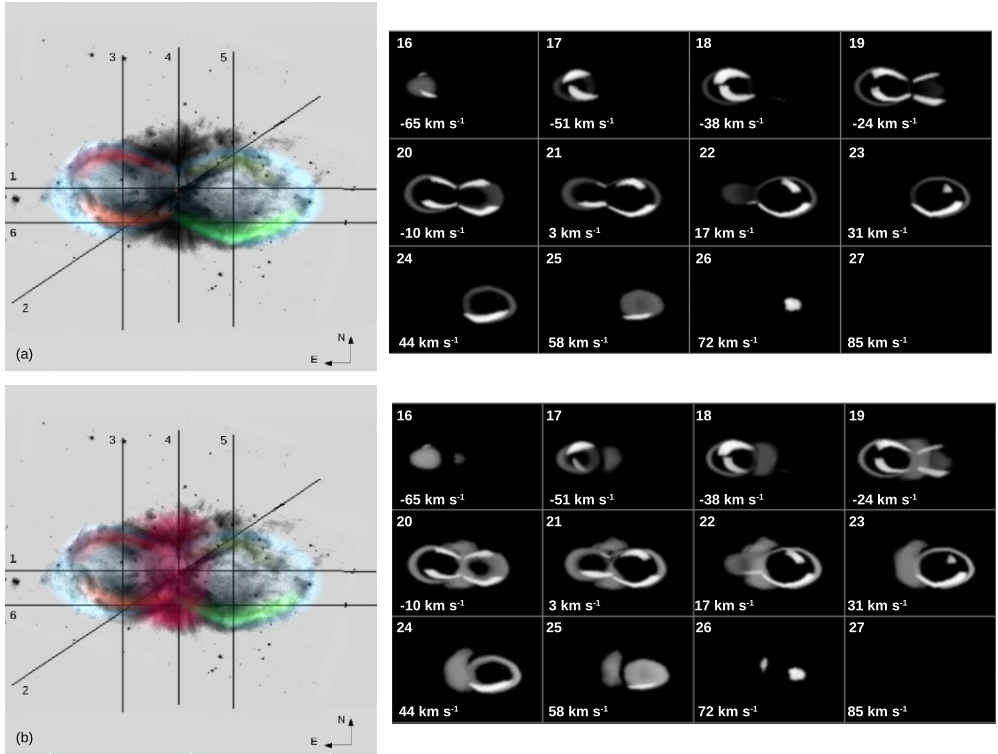

**Figure 9.** On the left we see the structures of the 3D MK model of NGC 2818 overlaid on the *HST* image, while on the right the 12 synthetic channel maps of these structures (numbers 16–27) deduced with SHAPE in [N II] λ6584 emission line are illustrated. The velocity range is $(-65, 85)$ km s$^{-1}$ with step 14 km s$^{-1}$, in proportion to the SAM-FP channel maps 16–27 of Fig. 6. In particular: (a) the BC along with the filamentary structures and the equivalent channel maps. (b) The final 3D MK model with all the components (lobe, filaments, and equatorial region) and the equivalent extracted channel maps. For further description, see Section 5.1.4.

## 6 DISCUSSION

NGC 2818 displays an overall bipolar morphology, with open-ended/broken bipolar lobes and a central region rich in several cometary knots and microstructures (Fig. 1).

The first 3D MK model of NGC 2818 by Vázquez (2012) was performed by considering a simple bipolar structure which adequately reproduced the PV of slit 1 of the obtained echelle spectra (Fig. 2). However, it was not adequate for reproducing the rest of the PV diagrams too (see the rest right panels in Fig. 2). Consequently, we needed to reconstruct an advanced and detailed 3D MK model, for which we combined echelle spectra and SAM-FP data. Eventually, apart from the BC, we ended in the 3D reconstruction of additional four filaments and the equatorial region (Fig. 7). This more complex model reproduces reasonably well all six PV diagrams (see Fig. 8). In addition, the synthetic channel maps of NGC 2818 presented in Fig. 9, are in sufficient agreement with the equivalent observational SAM-FP channel maps of Figs 5 and 6, both in the overall morphology and in the kinematics of the nebula. The successful reconstruction of both the echelle data and the channel maps, verifies the validity of the presented 3D MK model. Below, we discuss the basic attributes of NGC 2818, i.e. bipolarity and lobes fragmentation, filamentary structures, and cometary knots activity, based on our observational and modelling results. Furthermore, we present new detected structures in both SAM-FP and echelle data.

### 6.1 Bipolarity and lobes fragmentation in NGC 2818

For the justification of bipolarity in NGC 2818, the first scenario we adopt is based on the work of García-Segura, Taam & Ricker (2022), where the authors present a series of 2D HD simulations for the evolution of a proto-PNe to PNe with a binary system that has undergone a common envelope phase. The overall shape of NGC 2818 looks like the model C7-2.5 of fig. 3 in García-Segura et al. (2022), in which the dense and brighter outer shell relative to the central cavity displays an elliptical nebula. The central part displays a number of optically thick knots and filaments, being denser closer to the nucleus. The size of the outer shell becomes larger closer to the equator, similar to what we observe in NGC 2818. The snapshot of the model C7-2.5 in their fig. 3 corresponds to the computation at 10.000 yr. This is within the range of the kinematic age of NGC 2818 which, according to our model, was found between 6500–12300 yr (see Section 5.1.1). However, no fragmentation of the outer shell at the major axis is observed in their model, which is a typical attribute of NGC 2818.

The second scenario we adopt for the interpretation of bipolarity in NGC 2818 comes from Soker (2002), where a binary central system is also examined, with the additional parameter of jets/collimated fast winds ejection from the formed accretion disc. This scenario could also account for the characteristic fragmentation of the outer edges of the lobes of NGC 2818, as it is presented below.

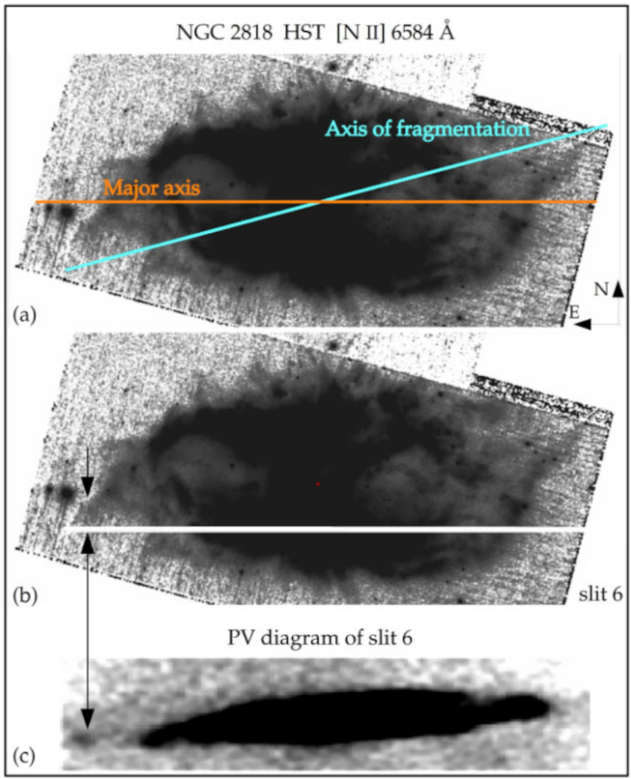

**Figure 10.** (a) The *HST* image of NGC 2818 in [N II] emission. The cyan line 'connects' the two diametrical fragmented regions of the two lobes of the nebula at a direction of 25 degrees with respect to the major axis of the nebula. (b) The same *HST* image as in (a), with the white line to represent the position of slit 6. (c) The PV diagram of slit 6. The black arrows mark the position of the small blob, which is evident in both imaging and echelle data. In the spectrum image, this blob seems to be detached with respect to the rest main structure of the PV area. We argue that this material is associated with the fragmented lobe. For further description, see Section 6.1.

In the *HST* image (Fig. 2a), it can be seen that both lobes are open-ended or fragmented, with the western part to be more fragmented with respect to the eastern part. Based on the aforementioned model of Soker (2002), if we assume that successive ejections of collimated fast outflows (jets) occur practically at the same direction during short time-scales, this would result in a bipolar morphology where a subsequent ejection takes place inside an already shaped lobe. Thus, we argue that these open-ended broken lobes could be the outcome of a subsequent strong expulsion of jets-structures which penetrated the former lobe. In Fig. 10(a), the cyan line represents the axis of fragmentation which passes through the central star of NGC 2818, and is tilted by ∼15 degrees with respect to the major axis of the main bipolar nebula (orange line). This slight misalignment could indicate a small wobbling of the system as well. Similar behaviour which includes 'wobbling' has been reported for the PN HB4 (Derlopa et al. 2019) and the PN Fleming 1 (Palmer et al. 1996; Boffin et al. 2012).

The ejection of material through these fragmented regions is also possible, as it is indicated in Figs 10(b) and (c). Fig. 10(b) presents the *HST* image along with the position of slit 6 (white line) which crosses the southern part of the nebula, while the corresponding PV diagram at low contrast is shown in Fig. 10(c). On the left part of the PV diagram, there is a very faint blob of radial velocity ∼−40 ± 20 km s$^{-1}$ marked with a black arrow, which seems not to be connected to the main 'body' of the spectrum. This blob is also present in the imaging data, but it is only evident with a lower contrast (see black arrow in Fig. 10b). We argue that this material is part of the fragmented eastern lobe which has now been separated from the overall bipolar structure. Unfortunately, this blob is not seen in the SAM-FP data, because it lies marginally out of their FOV.

The third scenario we propose for the morphology of NGC 2818, comes from the work presented by García-Segura & López (2000), where neither a binary system nor a rotating source is included, but involves the magnetic collimation axis of a single star by employing 3D MHD simulations. The authors suggest that, in order to achieve the formation of jets/ansae along with open-ended lobes, two basic factors must be taken into account: (a) the angle (the misalignment) between the magnetic collimation axis and the symmetric axis of the bipolar/elliptical wind outflow and (b) the mass-loss rate of the magnetized wind. From the presented models, a combination of the models B15/B45 and C15 of fig. 7 in García-Segura & López (2000) resembles quite well the overall morphology of NGC 2818 (for comparison with NGC 2818 see Fig. 11). The number of each model (e.g. B15) indicates the angle between the axis of the magnetic field and the axis of the bipolar outflow. The misalignment between the bipolar symmetric axis and the fragmentation axis of ∼15 degrees in NGC 2818 (see Fig. 10a), falls into the range of angles for the models of García-Segura & López (2000) that better reproduces the morphology of NGC 2818. It should be noted that, both B and C models produce shock-excited point-symmetric knots (see Fig. 11). The blob identified in slit 6 (Fig. 10c) of NGC 2818 may be the imprint of such jets activity.

A combination of the above presented models can account for the evolutionary history of NGC 2818, i.e. a binary central star evolved through a common envelope phase, in conjunction with jets activity and rotational motion, under the presence of a magnetic field, misaligned relative to the major bipolar axis of the nebula.

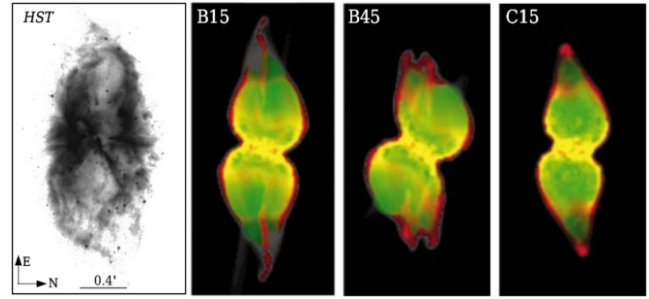

**Figure 11.** First panel: *HST* image of NGC 2818 in [N II] 6584 Å. Rest three panels: synthetic images for the B15, B45, and C15 models (García-Segura & López 2000). For further description, see Sections 6.1 and 6.2.

### 6.2 Filamentary structures and shock excitation in NGC 2818

Apart from the formation of jets and broken lobes mentioned in 6.1, the model of García-Segura & López (2000) also accounts for filamentary structures. Surprisingly, their B and C models clearly produce bright shock-heated filaments at the edges of the bipolar lobes (see the red-coloured structures in Fig. 11). Looking at the density distribution of the B45 model (right panels of fig. 4 in García-Segura & López 2000), one can see that at least two distinct components at different directions are formed and become evident after 1000 yr. This model resembles the overall structure of NGC 2818 seen in the *HST* image, with at least two visible filaments (see Fig. 11). Based on the above, we argue that the bright filaments observed in NGC 2818, and we included in our 3D MK model, may be associated with shock-excited material.

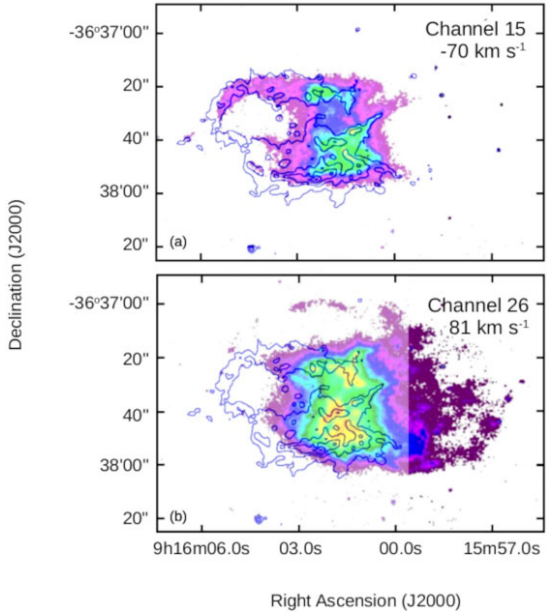

**Figure 12.** Two representative channel maps of the SAM-FP data in H$\alpha$ line in logarithmic scale, overlaid with the isocontours of the H$_2$ emission from Schild (1995). The spatial matching between the H$\alpha$ filament and H$_2$ emission is clearly deduced in the upper panel. The newly detected filament is illustrated in the lower panel, which is evident only in the SAM-FP data, and only in logarithmic scale. It is also noteworthy that the new filament is beyond the H$_2$ emission. For further analysis, see Sections 6.2, 6.3, and 6.4.

Shock excitation mechanism could account for the presence of strong H$_2$ emission in NGC 2818, which is evident in the north-eastern filaments and the southern part of the nebula (see Fig. 12) and, according to Schild (1995), is better explained by shock interaction (for centrally concentrated H$_2$ emission, see Section 6.3). Furthermore, regarding the southern part of the nebula of NGC 2818, Phillips & Cuesta (1998) also pointed out that the enhanced [N II]/H$\alpha$ and [S II]/H$\alpha$ line ratios in this area are linked with the shock-excited H$_2$ emission. The comparison of the [O III]/H$\alpha$ and [S II]/H$\alpha$ line ratios with the predictions from shock models suggested the presence of shocks with velocity higher than 100 km s$^{-1}$. Such high velocities have been found for both H$\alpha$ and [N II] lines in our SAM-FP data at the exact position where H$_2$ is detected, supporting the shock interaction scenario.

### 6.3 Cometary knots in NGC 2818

Another characteristic feature of NGC 2818 is the bright cometary knots, preferentially around the central star but not symmetrically distributed (Fig. 1). The central region of NGC 2818 resembles that of NGC 2440, although the latter is a case of a quadri-polar PN (i.e. a nebula consisted of two pairs of lobes), as its 3D MK model revealed (Lago & Costa 2016). Modelling the knots of NGC 2818 as individual components was not possible due to the low spatial resolution of the ground-based echelle spectroscopic data. Therefore, the central region of the nebula was modelled as a torus-like equatorial region, the geometrical locus where the cometary knots lie. Its P.A. was found to be 80 $\pm$ 3 deg, while its inclination 70 $\pm$ 3 deg, in agreement with the P.A./inclination angles of the main BC. Its expansion velocity was found to be lower compared to that of the BC, $V_{\rm exp} = 70 \pm 20$ km s$^{-1}$, which is consistent with the results reported for other bipolar PNe too (Banerjee et al. 1990; Meaburn et al. 1998 and references therein). SAM-FP data, also unveiled very faint, high velocity components in the central region (see Figs 5 and 6), likely associated with some knots seen in the *HST* image (Fig. 1). However, these high velocities were not observed in the echelle spectra.

For interpreting the origin of the cometary knots in NGC 2818, we take into account (i) the link between bright low-ionization lines ([N II] $\lambda$6584, [O I] $\lambda$6300) and H$_2$ S(1) emission, which has been verified in a number of PNe (Speck et al. 2002; Matsuura et al. 2009; Fang et al. 2015, 2018; Akras et al. 2017; Akras et al. 2020), and (ii) the fragmentation of the nebular equatorial region into knots and filaments which has been revealed through diffraction-limited H$_2$ images (e.g. Marquez-Lugo et al. 2013; Manchado et al. 2015). In particular for NGC 2818, there is a spatial coincidence of the [N II] and H$\alpha$ emission line distribution in the SAM-FP channel maps, with the distribution of the H$_2$ emission detected in NGC 2818 (Schild 1995). This is illustrated in Fig. 12(a) where a representative SAM-FP channel map in H$\alpha$ line is overlaid with the isocontours of the H$_2$ emission reported by Schild (1995). According to the author, the H$_2$ emission in the central part of the nebula – where there is also H$\alpha$ emission – is concentrated in small scale knots. UV-pumping was proposed as the dominant excitation mechanism of H$_2$ gas, while the H$_2$ number density of the knots should be $>10^4$ cm$^{-3}$ in order to shield the molecular gas from excitation, as it has been computed in PNe clumps (Meaburn et al. 1998, Matsuura et al. 2007, van Hoof et al. 2010). Given the strong central H$_2$ emission of NGC 2818 (Schild 1995), we could support that indeed these cometary knots result from the fragmentation of its molecular equatorial region, while a part of them has been excited through collisional excitation resulting in the emission of low-ionization lines as well. However, the low spatial resolution of the data does not allow the direct association of the cometary knots with H$_2$, as it has been reported in other cases of PNe (e.g. NGC 2346 in Manchado et al. 2015).

### 6.4 New detected north-eastern filament

The discovery of the new north-eastern filament in the SAM-FP data (Fig. 5) was an unforeseen result. This extremely faint and narrow filament is evident only in the H$\alpha$ line of the SAM-FP data, in a small range of velocities between 67 to 108 km s$^{-1}$ ($\pm$20 km s$^{-1}$) which corresponds to the channels maps 25–28 (see Fig. 5). As we can see in Fig. 12(b), its position is out of the region of H$_2$ emission detected in NGC 2818 by Schild (1995), and away from the main bipolar structure in general. At first glance, we considered the new filament as part of the eastern lobe of NGC 2818, due to its position and its curved shape. However, the eastern lobe is blue-shifted while the new filament is red-shifted, which indicates two different nebular components. Therefore, this new structure found in the SPM-FP data motivated us to examine more thoroughly the echelle data of the north-eastern part of the nebula and check if there was any very faint structure present, as it is described below.

Fig. 13 illustrates the *HST* image, the 26th SAM-FP channel map, and the PVs of slit 3 and slit 5. The red arc region in the SAM-FP channel map marks the position of the newly discovered filament in NGC 2818. The position of slit 3 is also presented and crosses the filament (vertical black line at the eastern part). Searching at the PV diagram of slit 3 for an equivalent structure, we ended up with the detection of a very faint red-shifted knotty structure only visible in the H$\alpha$ echelle data, as it is shown on the left in Fig. 13. The mean velocity of this structure is measured at 50 $\pm$ 20 km s$^{-1}$ in agreement with the velocity range of the new filament in the SAM-FP data, if we take into account the error of the SAM-FP instrument, which is of the order of 20 km s$^{-1}$.

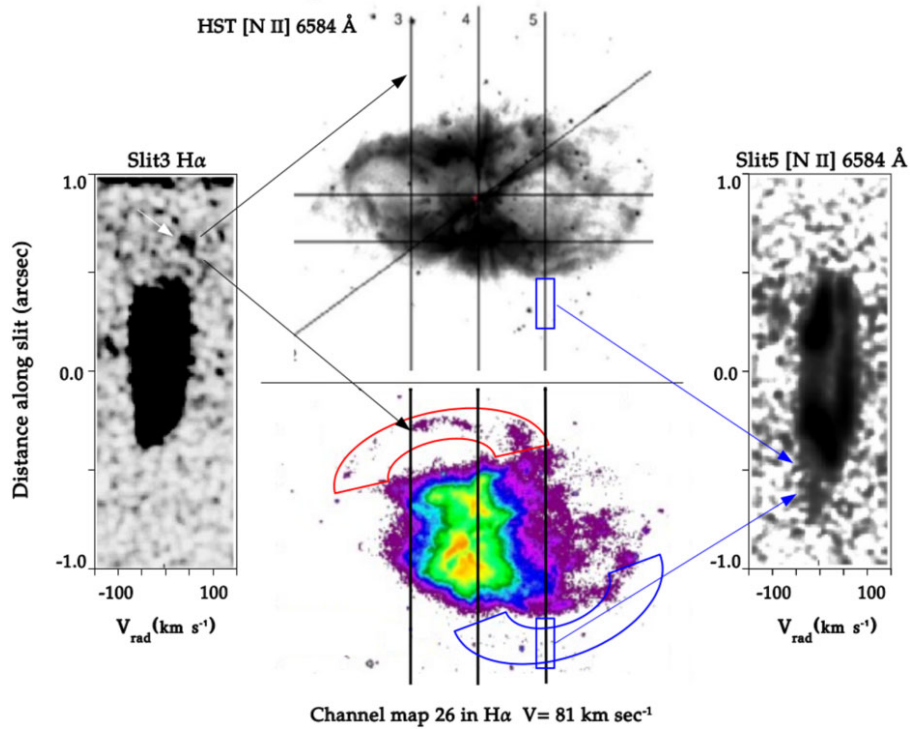

**Figure 13.** In this image, we illustrate assembled the more faint structures of our observational data. In the middle column, at the top is the *HST* image, while the bottom image shows the channel map 26 of the SAM-FP data in H$\alpha$ at velocity 81 km sec$^{-1}$. The three black vertical lines represent the positions of slits 3, 4, and 5. The PV diagrams from slits 3 in H$\alpha$ and slit 5 in [N II] lines are also presented in the left and right columns. The arrows mark the positions of the fainter and new detected structures. For further description, see Section 6.4.

Symmetry dictates the presence of an equivalent blue-shifted structure in the opposite diametric region of the new filament. Therefore, we searched for a potential structure in the region which is marked with the blue arc in the SPM-FP image of Fig. 13. Neither in the SPM-FP data nor in the *HST* image was something evident there, however it was evident in the PV diagram of slit 5 in [N II] 6584 Å, as it shown in the right part of Fig. 13. Indeed, apart from the main elliptical velocity structure in the PV of slit 5, we also found a faint elongated structure at its lower part, which is visible only in a high-contrast image. Its position in the SPM-FP and *HST* data is dictated with the two blue boxes and the equivalent two blue arrows (Fig. 13). This newly structure is blue-shifted with a mean velocity of $-40 \pm 20$ km s$^{-1}$ in agreement with what we were expecting. We argue that the main nebula of NGC 2818 is engulfed by another, larger and much fainter envelope/halo, probably created in a previous evolutionary stage of NGC 2818, as it has been reported for many cases of other PNe too (e.g. Corradi 2003).

## 7 CONCLUSIONS

A thorough analysis was conducted on the spectroscopic data of the PN NGC 2818. Long-slit echelle data were combined with newly Fabry–Perot data. These different types of observational data acted complementary for the reconstruction of a detailed 3D MK model of the nebula, and also for the confirmation of newly detected structures. The main conclusions drawn from this work are:

(i) According to the deduced 3D MK model, NGC 2818 is described by a bipolar shell component, four representative filamentary structures and an EC. All components have the same orientation, i.e. P.A. of $80 \pm 3$ deg and inclination of $70 \pm 3$ deg.

(ii) The model showed that the EC has the lower expansion velocity of the system, which is $70 \pm 20$ km s$^{-1}$. The velocity of the BC is $120 \pm 20$ km s$^{-1}$, while all the filamentary structures were found to expand at higher velocities at $180 \pm 20$ km s$^{-1}$.

(iii) Fabry–Perot data revealed a complex velocity field. The analysis of the emission lines showed regions with single peak profiles (indication of a single component), while others have double peak profiles (indication of more than one substructures).

(iv) A faint, blue-shifted structure/blob of $-40 \pm 20$ km s$^{-1}$ in the eastern part of the nebula was found in the echelle data, probably representing the imprint of previous jets ejection.

(v) A new filament at the north-eastern part of NGC 2818 has been detected in the SAM-FP data in the H$\alpha$ line, for a narrow velocity range between 67 and 108 km s$^{-1}$ ($\pm 20$ km s$^{-1}$). Part of this filament was then detected in the echelle spectra too, with an average velocity of $50 \pm 20$ km s$^{-1}$, indicating a much larger and fainter envelope – probably of former evolutionary stage – that now surrounds the nebula.

(vi) A faint, blue-shifted sub-structure in the echelle data has been detected at the south-western region of the nebula, at a velocity of $-40 \pm 20$ km s$^{-1}$. We argue that this structure is part of the equivalent symmetric structure of the new detected north-eastern filament.

(vii) An angle of $\sim 15$ degrees between the major axis of the nebula and the axis of the fragmented bipolar lobes was measured. This can be related to performed MHD simulations, according to which a misalignment between the axis of a magnetized jet with respect to the symmetry axis of the bipolar/elliptical shell, can be responsible for asymmetric fragmentation, as is the case of NGC 2818.

(viii) The overall structure of NGC 2818 can be reconstructed by HD simulations as the result of the evolution of a binary system that underwent the common envelope phase, in conjunction with the ejections of a magnetized jet, misaligned with respect to the symmetry axis of the bipolar/elliptical shell.

**ACKNOWLEDGEMENTS**

SA acknowledges support under the grant 5077 financed by IAASARS/NOA. The authors would like to thank the anonymous referee whose recommendations helped us to significantly improve this paper, and also Dr Wolfgang Steffen for his valuable comments on this project. Based on observations obtained at the Southern Astrophysical Research (SOAR) telescope, which is a joint project of the Ministerio da Ciência, Tecnologia, e Inovação (MCTI) da Republica Federativa do Brasil, the U.S.A. National Optical Astronomy Observatory (NOAO), the University of North Carolina at Chapel Hill (UNC), and Michigan State University (MSU).

**DATA AVAILABILITY**

The data and results of this article will be shared on reasonable request to the corresponding author.

**REFERENCES**

Akras S., Gonçalves D. R., 2016, MNRAS, 455, 930
Akras S., López J. A., 2012, MNRAS, 425, 2197
Akras S., Steffen W., 2012, MNRAS, 423, 925
Akras S., Boumis P., Meaburn J., Alikakos J., López J. A., Gonçalves D. R., 2015, MNRAS, 452, 2911
Akras S., Clyne N., Boumis P., Monteiro H., Gonçalves D. R., Redman M. P., Williams S., 2016, MNRAS, 457, 3409
Akras S., Gonçalves D. R., Ramos-Larios G., 2017, MNRAS, 465, 1289
Akras S., Gonçalves D. R., Ramos-Larios G., Aleman I., 2020, MNRAS, 493, 3800
Bailer-Jones C. A. L., Rybizki J., Fouesneau M., Mantelet G., Andrae R., 2018, AJ, 156, 58
Bailer-Jones C. A. L., Rybizki J., Fouesneau M., Demleitner M., Andrae R., 2021, AJ, 161, 147
Balick B., 1987, AJ, 94, 671
Balick B., Frank A., 2002, ARA&A, 40, 439
Balick B., Frank A., Liu B., Corradi R., 2018, ApJ, 853, 168
Bandyopadhyay R., Das R., Mondal S., Ghosh S., 2020, MNRAS, 496, 814
Banerjee D. P. K., Anandarao B. G., Jain S. K., Mallik D. C. V., 1990, A&A, 240, 137
Boffin H. M. J., Miszalski B., Rauch T., Jones D., Corradi R. L. M., Napiwotzki R., Day-Jones A. C., Köppen J., 2012, Science, 338, 773
Chong S. N., Kwok S., Imai H., Tafoya D., Chibueze J., 2012, ApJ, 760, 115
Clark D. M., López J. A., Steffen W., Richer M. G., 2013, AJ, 145, 57
Clyne N., Akras S., Steffen W., Redman M. P., Gonçalves D. R., Harvey E., 2015, A&A, 582, A60
Corradi R. L. M., 2003, in Kwok S., Dopita M., Sutherland R., eds, Proc. IAU Symp. 209, Planetary Nebulae: Their Evolution and Role in the Universe. Astron. Soc. Pac., San Francisco, p. 447
Danehkar A., 2022, ApJS, 260, 14
Dayal A., Sahai R., Watson A. M., Trauger J. T., Burrows C. J., Stapelfeldt K. R., Gallagher John S. I., 2000, AJ, 119, 315
De Marco O. et al., 2022, Nat. Astron., 6, 1421
de la Fuente E., Trinidad M. A., Tafoya D., Toledano-Juárez I., García-Flores S., 2022, PASJ, 74, 594
Derlopa S., Akras S., Boumis P., Steffen W., 2019, MNRAS, 484, 3746
Derlopa S., Boumis P., Chiotellis A., Steffen W., Akras S., 2020, MNRAS, 499, 5410
Dufour R. J., 1984, ApJ, 287, 341
Fang X., Guerrero M. A., Miranda L. F., Riera A., Velázquez P. F., Raga A. C., 2015, MNRAS, 452, 2445
Fang X., Zhang Y., Kwok S., Hsia C.-H., Chau W., Ramos-Larios G., Guerrero M. A., 2018, ApJ, 859, 92
Gaia Collaboration, 2018, A&A, 616, A1
Gaia Collaboration, 2021, A&A, 649, A1
García-Segura G., López J. A., 2000, ApJ, 544, 336
García-Segura G., Taam R. E., Ricker P. M., 2022, MNRAS, 517, 3822
Gómez-Gordillo S., Akras S., Gonçalves D. R., Steffen W., 2020, MNRAS, 492, 4097
Gonçalves D. R., Corradi R. L. M., Mampaso A., 2001, ApJ, 547, 302
Górny S. K., Schwarz H. E., Corradi R. L. M., Van Winckel H., 1999, A&AS, 136, 145
Guerrero M. A., Cazzoli S., Rechy-García J. S., Ramos-Larios G., Montoro-Molina B., Gómez-González V. M. A., Toalá J. A., Fang X., 2021, ApJ, 909, 44
Harvey E., Redman M. P., Boumis P., Akras S., 2016, A&A, 595, A64
Harvey E. J. et al., 2020, MNRAS, 499, 2959
Hora J. L., Latter W. B., Allen L. E., Marengo M., Deutsch L. K., Pipher J. L., 2004, ApJS, 154, 296
Hora J. L., Latter W. B., Smith H. A., Marengo M., 2006, ApJ, 652, 426
Jones D., Boffin H. M. J., 2017, Nat. Astron., 1, 0117
Kahn F. D., West K. A., 1985, MNRAS, 212, 837
Kamiński T., Steffen W., Bujarrabal V., Tylenda R., Menten K. M., Hajduk M., 2021, A&A, 646, A1
Kastner J. H., Moraga Baez P., Balick B., Bublitz J., Montez R., Frank A., Blackman E., 2022, ApJ, 927, 100
Kwok S., 2000, The Origin and Evolution of Planetary Nebulae. Cambridge Univ. Press, Cambridge, p. 33
Kwok S., Purton C. R., Fitzgerald P. M., 1978, ApJ, 219, L125
Lago P. J. A., Costa R. D. D., 2016, Rev. Mex. Astron. Astrofis., 52, 329
López J.-A., Richer M. G., García-Díaz M.-T., Clark D. M., Meaburn J., Riesgo H., Steffen W., Lloyd M., 2012, Revista Mexicana de Astronomía y Astrofísica, 48, 3
Manchado A., Stanghellini L., Villaver E., García-Segura G., Shaw R. A., García-Hernández D. A., 2015, ApJ, 808, 115
Mari M. B., Gonçalves D. R., Akras S., 2023, MNRAS, 518, 3908
Marquez-Lugo R. A., Ramos-Larios G., Guerrero M. A., Vázquez R., 2013, MNRAS, 429, 973
Mata H. et al., 2016, MNRAS, 459, 841
Matsuura M. et al., 2007, MNRAS, 382, 1447
Matsuura M. et al., 2009, ApJ, 700, 1067
Meaburn J., Boumis P., 2010, MNRAS, 402, 381
Meaburn J., Clayton C. A., Bryce M., Walsh J. R., Holloway A. J., Steffen W., 1998, MNRAS, 294, 201
Meaburn J., López J. A., Gutiérrez L., Quiróz F., Murillo J. M., Valdéz J., Pedrayez M., 2003, Rev. Mex. Astron. Astrofis., 39, 185
Meaburn J., Boumis P., López J. A., Harman D. J., Bryce M., Redman M. P., Mavromatakis F., 2005, MNRAS, 360, 963
Meatheringham S. J., Wood P. R., Faulkner D. J., 1988, ApJ, 334, 862
Mendes de Oliveira C., Amram P., Quint B. C., Torres-Flores S., Barbá R., Andrade D., 2017, MNRAS, 469, 3424
Mermilliod J. C., Clariá J. J., Andersen J., Piatti A. E., Mayor M., 2001, A&A, 375, 30
Moraga Baez P., Kastner J. H., Balick B., Montez R., Bublitz J., 2023, ApJ, 942, 15
Paczynski B., 1976, in Eggleton P., Mitton S., Whelan J., eds, Proc. IAU Symp. 73, Structure and Evolution of Close Binary Systems. Reidel, Dordrecht, p. 75
Palmer J. W., Lopez J. A., Meaburn J., Lloyd H. M., 1996, A&A, 307, 225
Peimbert M., Torres-Peimbert S., 1983, in IAU Symp., Vol. 103, p. held at University College, London, U.K. August 9-13, 1982 Ed. by D.R. Flower., 233-242
Phillips J. P., 2004, MNRAS, 353, 589
Phillips J. P., Cuesta L., 1998, A&AS, 133, 381
Phillips J. P., Ramos-Larios G., 2010, MNRAS, 405, 2179
Rodríguez-González J. B. et al., 2021, MNRAS, 501, 3605
Ruiz N., Guerrero M. A., Chu Y.-H., Gruendl R. A., 2011, AJ, 142, 91

Sabin L. et al., 2017, MNRAS, 467, 3056
Sahai R., Morris M. R., Villar G. G., 2011, AJ, 141, 134
Schild H., 1995, A&A, 297, 246
Shklovsky I. S., 1956, AZh, 33, 315
Soker N., 2002, ApJ, 568, 726
Soker N., Rappaport S., 2000, ApJ, 538, 241
Speck A. K., Meixner M., Fong D., McCullough P. R., Moser D. E., Ueta T., 2002, AJ, 123, 346
Steffen W., Koning N., Wenger S., Morisset C., Magnor M., 2011, Proc. IEEE, 17, 454
Tifft W. G., Connolly L. P., Webb D. F., 1972, MNRAS, 158, 47
Toalá J. A., Guerrero M. A., Bianchi L., Chu Y. H., De Marco O., 2020, MNRAS, 494, 3784
Tokovinin A., Tighe R., Schurter P., Cantarutti R., van der Bliek N., Martinez M., Mondaca E., Montane A., 2008, in Hubin N., Max C. E., Wizinowich P. L., eds, Proc. SPIE Conf. Ser. Vol. 7015, Adaptive Optics Systems. SPIE, Bellingham, p. 70154C
Vázquez R., 2012, ApJ, 751, 116
Wesson R. et al., 2024, MNRAS, 528, 3392
Zhang C. Y., 1995, ApJS, 98, 659
van Hoof P. A. M. et al., 2010, A&A, 518, L137
van de Steene G. C., Zijlstra A. A., 1995, A&A, 293, 541

This paper has been typeset from a T$_E$X/L$^A$T$_E$X file prepared by the author.